\documentclass[letter,11pt]{article}

\usepackage{newlfont}
\usepackage[small]{caption2}
\usepackage[below]{placeins}
\usepackage{flafter}
\usepackage{amsfonts}
\usepackage{amsmath}
\usepackage{amssymb}
\usepackage[american]{babel}
\usepackage{graphicx}
\usepackage{subfigure}
\usepackage{multirow}
\newcommand{\ket}[1]{\big| #1 \big\rangle}
\newcommand{\bra}[1]{\big\langle #1 \big|}
\newcommand{\braket}[2]{\left\langle #1 \big| #2 \right\rangle}                 

\begin{document}


\title{Quantum Hitting Time on the Complete Graph}
\author{R.A.M. Santos and R.~Portugal \\
\\
{\small Laborat\'{o}rio Nacional de Computa\c{c}\~{a}o Cient\'{\i}fica - LNCC} \\
{\small Av. Get\'{u}lio Vargas 333, Petr\'{o}polis, RJ, 25651-075, Brazil}
}


\date{\today}
\maketitle

\begin{abstract}
Quantum walks play an important role in the area of quantum algorithms. Many interesting problems can be reduced to searching marked states in a quantum Markov chain. In this context, the notion of quantum hitting time is very important, because it quantifies the running time of the algorithms. Markov chain-based algorithms are probabilistic, therefore the calculation of the success probability is also required in the analysis of the computational complexity. Using Szegedy's definition of quantum hitting time, which is a natural extension of the definition of the classical hitting time, we present analytical expressions for the hitting time and success probability of the quantum walk on the complete graph.
\end{abstract}

\section{Introduction}

The notion of hitting time in classical Markov chains plays an important role in Computer Science. The hitting time is used in Monte Carlo algorithms, and in randomized algorithms in general, as the running time to find a solution~\cite{MR95}. Expressions for the classical hitting time were calculated analytically for many graphs~\cite{Aldous}.

It is not straightforward to generalize the classical definition of hitting time to the quantum realm. Kempe~\cite{Kempe} has provided two definitions and proved that a quantum walker hits the opposite corner of a $n$-hypercube in time $O(n)$. Krovi and Brun~\cite{KB06} have provided a definition of average hitting time that requires a partial measurement of the position of the walker at each step. Kempf and Portugal~\cite{KP09} have discussed the relation between hitting times and the walker's group velocity.

Inspired on Ambainis' algorithm~\cite{Amb03} for solving the element distinctness problem, Szegedy~\cite{Sze04} was able to abstract out the mathematical structure of that algorithm and to provide a definition of quantum hitting time, that is a natural generalization of the classical definition of hitting time. Years of effort show that the establishment of that definition is far from trivial. Recently, Magniez \textit{et al.}~\cite{MNRS} have extended Szegedy's work to non-symmetric ergodic Markov chains and have improved the probability to find a marked state using Tulsi's method~\cite{Tulsi08}.

In this work we calculate analytically Szegedy's  hitting time and the probability of finding a set of marked vertices on the complete graph. This calculation clarifies many points of Szegedy's definition, such as the analytical behavior of the time average of the quantity $\big\|\,U^t\ket{\phi_0}-\ket{\phi_0}\,\big\|$, where $\ket{\phi_0}$ is the initial condition and $U^t$ is the evolution operator after $t$ steps. We show why the calculation of the hitting time is easier than the calculation of the success probability. The eigenspace associated with the eigenvalue 1 of the evolution operator plays no role in the calculation of the hitting time, but must be taken into account in the calculation of the success probability.

The paper is organized as follows. In Sec.~\ref{ht_sec_1} we review the basic operators of a bipartite graph that are needed in the definition of the evolution operator. In Sec.~\ref{ht_sec_U} we review Szegedy's definition of the quantum walk's evolution operator and the method to obtain part of its spectral decomposition. In Sec.~\ref{ht_sec_HT} we review Szegedy's definition of hitting time. In Sec.~\ref{ht_sec_CG} we calculate the hitting time and the probability of finding a marked vertex on the complete graph.

\section{Reflection operators in a bipartite graph}\label{ht_sec_1}

In order to define the quantum hitting time in a graph, Szegedy~\cite{Sze04} has proposed a quantum walk driven by reflection operators in an associated bipartite graph obtained from the original one by a process of duplication, as explained in Sec.~\ref{ht_sec_HT}.

Consider a bipartite graph between the set of vertices $ X $ and $ Y $ of same cardinality. Denote by $ x $ and $ y $ generic vertices in sets $ X $ and $ Y $. The stochastic matrices $P$ and $Q$ associated with this graph are defined such that $p_{x y}$ is the inverse of the outdegree of the vertex $ x $, if there is a directed edge from $x$ to $y$, otherwise $p_{x y}=0$. Analogously, $q_{yx}$ is either the inverse of the outdegree of the vertex $y$ or zero. The variables $ p_{x y} $ and $q_{y x}$ satisfy
\begin{eqnarray}
  \sum_{y\in Y} p_{x y}=1 && \forall x \in X,  \label{ht_pxy}\\
  \sum_{x\in X} q_{y x}=1 && \forall y \in Y.   \label{ht_qyx}
\end{eqnarray}
To define a quantum walk in the bipartite graph, we associate with the graph a Hilbert space ${\cal H}^{n^2} = {\cal H}^{n}\otimes {\cal H}^{n} $, where $ n = | X | = | Y | $. The computational basis of the first component is $ \big \{\ket {x}: x \in X \big \} $ and of the second $ \big \{\ket{y}: y \in Y \big \} $. The computational basis of $ {\cal H}^{n^2} $ is $ \big \{\ket {x, y}: x \in X, y \in Y \big \} $. In the quantum case, instead of using the stochastic matrices $ P $ and $ Q $ of the classical random walk, we define the operators $ A: {\cal H}^n \rightarrow {\cal H}^{n^2} $ and $ B: {\cal H}^n \rightarrow {\cal H}^{n^2} $ as follows
\begin{eqnarray}
  A &=& \sum_{x\in X} \ket{\alpha_x}\bra{x}, \label{ht_A}\\
  B &=& \sum_{y\in Y} \ket{\beta_y}\bra{y}, \label{ht_B}
\end{eqnarray}
where
\begin{eqnarray}
  \ket{\alpha_x} &=& \ket{x}\otimes \left(\sum_{y\in Y} \sqrt{p_{x y}} \, \ket{y}\right), \label{ht_alpha_x} \\
  \ket{\beta_y}  &=&  \left(\sum_{x\in X} \sqrt{q_{y x}} \, \ket{x}\right)\otimes \ket{y}. \label{ht_beta_y}
\end{eqnarray}
$ A $ and $ B $ are $ n^2 \times n $ matrices. Eqs.~(\ref{ht_A}) and (\ref{ht_B}) tell us that the columns of $A$ are the vectors $\ket{\alpha_x}$ and the columns of $B$ are $\ket{\beta_y}$. Vectors $\ket{\alpha_x}$ and $\ket{\beta_y}$ obey
\begin{eqnarray}
  \braket{\alpha_x}{\alpha_{x^\prime}} &=& \delta_{x,x^\prime}, \\
  \braket{\beta_y}{\beta_{y^\prime}} &=& \delta_{y,y^\prime}.
\end{eqnarray}
Therefore
\begin{eqnarray}
  A^T A &=& I_n, \label{ht_ATA}\\
  B^T B &=& I_n. \label{ht_BTA}
\end{eqnarray}
These equations imply that $ A $ and $ B $ preserve the norm of vectors, so if $ \ket{\mu} $ is a unit vector of $ {\cal H}^n $, then $ A \ket{\mu} $ is a unit vector of $ {\cal H} ^{n^2} $. The same for $ B $.

Of course we will investigate the product in the reverse order. Using Eqs.~(\ref{ht_A}) and (\ref{ht_B}) we obtain
\begin{eqnarray}
  A A^T &=& \sum_{x\in X} \ket{\alpha_x}\bra{\alpha_x}, \label{ht_AAT}\\
  B B^T &=& \sum_{y\in Y} \ket{\beta_y}\bra{\beta_y}. \label{ht_BBT}
\end{eqnarray}
Using Eqs.~(\ref{ht_ATA}) and (\ref{ht_BTA}) we have $ (AA^T)^2 = AA^T $ and $ (BB^T)^2 = BB^T $. So let us define the projectors
\
\begin{eqnarray}
  \Pi_A &=& A A^T, \\
  \Pi_B &=& B B^T.
\end{eqnarray}
Eqs.~(\ref{ht_AAT}) and (\ref{ht_BBT}) show that $\Pi_A$ project a generic vector of $ {\cal H} ^{n^2} $ to the subspace $ {\cal H}_A $ spanned by $\big\{\ket{\alpha_x}: x\in X\big\}$ and $\Pi_B$ to the subspace ${\cal H}_B $ spanned by $\big\{\ket{\beta_y}: y\in Y\big\}$.

We can now define the reflection operators associated with each of these projectors
\begin{eqnarray}
  \cal R_A &=& 2\,\Pi_A - I_{n^2}, \label{ht_RA}\\
  \cal R_B &=& 2\,\Pi_B - I_{n^2}. \label{ht_RB}
\end{eqnarray}
$\cal R_A$ reflects a generic vector in ${\cal H}^{n^2}$ around ${\cal H}_A$ and $\cal R_B$ around ${\cal H}_B$.

Now it is time to establish a connection between the subspaces $ {\cal H}_A $ and $ {\cal H}_B $. The best choice is to analyze the angles between the set of vectors $\big\{\ket{\alpha_x}: x\in X\big\}$ with $\big\{\ket{\beta_y}: y\in Y\big\}$. Let us define the matrix of inner products $ C $ such that $C_{x y}=\braket{\alpha_x}{\beta_y}$. Using Eqs.~(\ref{ht_alpha_x}) and (\ref{ht_beta_y}) we can express the components of $ C $ in terms of transition probabilities as $C_{x y}=\sqrt{p_{x y} q_{y x}}$, and in matrix form
\begin{eqnarray}\label{ht_C}
    C &=& A^T B.
\end{eqnarray}
 $ C $ is a square matrix of dimension $ n $. It provides essential information on the quantum walk that will be defined on the bipartite graph. $ C $ is not a normal operator in general. Its singular values and vectors play an important role in the dynamics of the quantum walk.


The theorem of singular value decomposition~\cite{NC00} states that there are unitary matrices $U$ and $V$  such that
\begin{equation}\label{ht_C_2}
    C=U D V^\dagger,
\end{equation}
where $D$ is a diagonal matrix of dimension $n$ with nonnegative real components. The diagonal elements are called singular values and univocally determined. Matrices $U$ and $V$ can be determined through the application of the spectral theorem to $C^\dagger C$, which is a semidefined positive matrix.

Let $\ket{\nu_j}$ and $\ket{\mu_j}$ be the right and left singular vectors respectively and $\lambda_j$ the corresponding singular values, then
\begin{eqnarray}
  C\ket{\nu_j} &=& \lambda_j \, \ket{\mu_j}, \label{ht_C_nu_i} \\
  C^T\ket{\mu_j} &=& \lambda_j \, \ket{\nu_j}. \label{ht_CT_mu_i}
\end{eqnarray}
Multiplying Eq.~(\ref{ht_C_nu_i}) by $A$ and Eq.~(\ref{ht_CT_mu_i}) by $B$ we obtain
\begin{eqnarray}
  \Pi_A\,B\ket{\nu_j} &=& \lambda_j \, A\ket{\mu_j}, \label{ht_Pi_A_nu_i} \\
  \Pi_B\, A\ket{\mu_j} &=& \lambda_j \, B\ket{\nu_j}. \label{ht_Pi_B_mu_i}
\end{eqnarray}
The action of operators $ A $ and $ B $ preserves the norm of vectors, then vectors $A\ket{\mu_j}$ and $B\ket{\nu_j}$  are unitary. Projectors either decrease the norm of vectors or maintain invariant. Using Eq.~(\ref{ht_Pi_A_nu_i}) we conclude that the singular values satisfy the inequalities $0\leq \lambda_j \leq 1$. So, we can define $\theta_j$ such that $\lambda_j=\cos \theta_j$, where $0\leq \theta_j\leq \pi/2$. The geometric interpretation of $ \theta_j $ is the angle between the vectors $A\ket{\mu_j}$ and $B\ket{\nu_j}$, that can be confirmed by using Eqs.~(\ref{ht_C}) and (\ref{ht_C_nu_i}).

\section{Evolution Operator and its Spectral Decomposition}\label{ht_sec_U}

Let us consider a bipartite graph such that $X=Y$, $P=P^T$ and $P=Q$. Szegedy~\cite{Sze04} has defined the one-step evolution operator in the Hilbert space of this graph as
\begin{equation}\label{ht_U_ev}
    U_P := \cal R_B \, \cal R_A,
\end{equation}
where $\cal R_A$ and $\cal R_B$ are given by Eqs.~(\ref{ht_RA}) and (\ref{ht_RB}).

Eqs.~(\ref{ht_Pi_A_nu_i}) and (\ref{ht_Pi_B_mu_i}) show that the projectors $ \Pi_A $ and $ \Pi_B $  have a symmetric action over vectors $ A \ket{\mu_j} $ and $ B \ket{\nu_j} $ for each $ j $. It is expected that the action of the reflection operators $ \cal R_A $ and $ \cal R_B $ on a linear combination of $ A \ket{\mu_j}$ and $ B \ket{\nu_j}$  results in a vector in the plane spanned by $ A \ket{\mu_j}$ and $ B \ket{\nu_j}.$ That is, this plane is invariant under the action of $ U_P $. So let us try the following \textit{Ansatz} for the eigenvectors of $ U_P $
\begin{equation}\label{ht_U A_B}
    U_P\big(a\,A\ket{\mu_j}+b\,B\ket{\nu_j}\big) = \lambda_j^\prime \big(a\,A\ket{\mu_j}+b\,B\ket{\nu_j}\big).
\end{equation}
The goal is to find $a$, $b$ and $\lambda_j^\prime$ that obey Eq.~(\ref{ht_U A_B}). Using definition (\ref{ht_U_ev}) for $U_P$, we eventually obtain that the vectors
\begin{equation}\label{ht_alpha_i}
    \ket{\alpha_j^\pm} = \frac{A\ket{\mu_j} - {\textrm e}^{\pm\,i\theta_j} B\ket{\nu_j}}{\sqrt 2 \sin \theta_j}
\end{equation}
are normalized eigenvectors with eigenvalues $\textrm{e}^{\pm 2 i\theta_j}$ when $0<\theta_j \leq \pi/2$. We have obtained at most $2n$ eigenvectors of $U_P$ so far, because $C$ has dimension $n$. In fact, the exact number depends on the multiplicity of the singular value $1$. For $\theta_j=0$, $A\ket{\mu_j}$ and $B\ket{\nu_j}$ do not span a two dimensional subspace, because they are colinear. Let us consider vectors $A\ket{\mu_j}$. From Eqs.~(\ref{ht_Pi_A_nu_i}) and (\ref{ht_Pi_B_mu_i}) we verify that they are invariant under the action of $\Pi_A$ and $\Pi_B$. Then, they are invariant under the action of $\cal R_A$ and $\cal R_B$. Then $A\ket{\mu_j}$ are eigenvectors of $U_P$ with eigenvalue 1. If the multiplicity of the singular value $1$ is $k$, then we have obtained $2n-k$ eigenvectors of $U_P$ so far. The remaining $n^2-2n+k$ eigenvectors cannot be found by using the singular values and vectors of matrix $C$, on the other hand, it is straightforward to show that the missing ones have eigenvalue 1.

\section{Quantum Hitting Time}\label{ht_sec_HT}

Szegedy~\cite{Sze04} has defined a notion of quantum hitting time that is a natural generalization of the concept of classical hitting time. Let $\Gamma(X,E)$ be a connected, undirected and non-bipartite graph, where $X$ is the set of vertices and $E$ is the set of edges. Define a bipartite graph associated with $ \Gamma (X, E) $ through a process of duplication. $ X $ and $ Y $ are the sets of vertices of same cardinality of the bipartite graph. Each edge $ \{x_i, x_j\} $ in $ E $ of the original graph $\Gamma(X,E)$ is converted into two edges in the bipartite graph $\{x_i,y_j\}$ and $\{y_i,x_j\}$.

The quantum walk on the bipartite graph is defined by the evolution operator $ U_P $ given by Eq.~(\ref{ht_U_ev}). In the bipartite graph, an application of $ U_P $ corresponds to two quantum steps of the walk, from $X$ to $Y$ and from $Y$ to $X$. We have to take the partial trace over the space associated with $ Y $ to get the state on the set $ X $.

In the \textit{classical case}, the {hitting time} $H_{x_0\,x_f}$ is the expected number of steps in a random walk that starts at $x_0$ and ends upon first reaching $x_f$~\cite{MR95}. This definition can be generalized to what is called average hitting time. Instead of departing from vertex $x_0$, the initial vertex can be sampled according to a probability distribution $\sigma$, such that $\sum_{x\in X} \sigma(x)=1$. Also, instead of reaching vertex $x_f$, one may consider the case of reaching a subset $M$ of $X$. So, the hitting time $H_{\sigma\,M}$ is the expected number of steps in a random walk that starts at a vertex that is sampled according to a probability distribution $\sigma$ and ends upon first reaching any vertex of $M$. Szegedy's definition is the quantum analogue of that last version. It is at least quadratically faster than the classical case.

To define the \textit{quantum hitting time}, Szegedy has used a modified evolution operator $ U_{P^\prime} $ associated with a modified directed bipartite graph obtained in the following form. Each edge of an undirected graph can be viewed as two opposite directed edges, since directed edges are fused to form the non-directed edge. The modified directed graph is the bipartite graph obtained by removing all directed edges leaving the vertices of the set $ M $, but keeping the directed edges that are arriving. This means that if the walker reaches a marked vertex, it will be stuck in that vertex in the following steps. To calculate the classical hitting time, the original and the modified bipartite graphs are equivalent. However, since the stochastic matrix has been modified, in the quantum case the evolution operator $ U_{P^\prime} $ is different from $ U_P $. If the walk starts uniformly distributed, the modulus of the amplitude probabilities at the marked vertices will increase at some specific moments. The modified stochastic matrix $P^\prime$ is given by
\begin{equation}\label{ht_pprime}
    p_{x y}^\prime = \left\{
                       \begin{array}{ll}
                         p_{x y}, & \hbox{$x\not\in M$;} \\
                         \delta_{x y}, & \hbox{$x\in M$.}
                       \end{array}
                     \right.
\end{equation}

The initial condition of the quantum walk is
\begin{equation}
    \ket{\psi(0)} = \frac{1}{\sqrt n} \sum_{
\begin{subarray}{c}
{x\in X}\\
y\in Y
\end{subarray}
} \sqrt{p_{xy}} \ket{x,y}.
\end{equation}
Note that $\ket{\psi(0)}$ is an eigenvector of $ U_P $ with eigenvalue $1$, when the probability distribution $ p_{xy}$ is symmetric. However, $\ket{\psi(0)}$ is not an eigenvector of $ U_{P^\prime} $ in general. Before describing the evolution of the quantum walk driven by the modified operator $ U_{P^\prime} $, let us define the \textit{quantum hitting time}~\cite{Sze04}.

\

\noindent\textbf{Definition}  The \textit{quantum hitting time} $H_{P,M}$ of a quantum walk with evolution operator $U_P$ given by Eq.~(\ref{ht_U_ev}) and initial condition $\ket{\psi(0)}$ is defined as the least number of steps $ T $ such that
\begin{equation}
    F(T) \geq 1-\frac{m}{n},
\end{equation}
where $ m $ is the number of marked vertices, $ n $ is the number of vertices of the original graph
and $F(T)$ is
\begin{equation}\label{ht_D_T}
    F(T) = \frac{1}{T+1}\sum_{t=0}^T \Big\|\ket{\psi(t)}-\ket{\psi(0)}\Big\|^2,
\end{equation}
where $\ket{\psi(t)}=U_{P^\prime}^t\ket{\psi(0)}$ and $U_{P^\prime}^t$ is the evolution operator after $t$ steps using the modified stochastic matrix.


\

Only the singular values of $C$ that are different from 1 are used in the calculation of hitting time. To make this point clear, let us write the initial condition in the eigenbasis of the evolution operator
\begin{equation}\label{ht_psi_0_generic}
    \ket{\psi(0)} = \sum_{j=1}^{n-k}\,\big(c_j^+\ket{\alpha_j^{+}} + c_j^- \ket{\alpha_j^{-}}\big) + \sum_{j=n-k+1}^{n^2-n+k}\,c_j\ket{\alpha_j},
\end{equation}
where $k$ is the multiplicity of the singular value 1. The coefficients $c_j^\pm$ are given by
\begin{equation}\label{ht_c_j_pm}
    c_j^\pm = \braket{\alpha_{j}^\pm}{\psi(0)}
\end{equation}
and obey the constraint
\begin{equation}\label{ht_cpm_c_j}
   \sum_{j=1}^{n-k}\,\left(\big|c_j^+\big|^2 + \big|c_j^-\big|^2\right) + \sum_{j=n-k+1}^{n^2-n+k} \, \big|c_j \big|^2=1.
\end{equation}
Applying $U_{P^\prime}^t$ to $ \ket{\psi(0)}$ we obtain
\begin{equation}\label{ht_psi_t_generic}
    \ket{\psi(t)} = \sum_{j=1}^{n-k}\,\left(c_j^+\textrm{e}^{ 2 i\theta_j t}\ket{\alpha_j^{+}} + c_j^- \textrm{e}^{- 2 i\theta_j t}\ket{\alpha_j^{-}}\right) + \sum_{j=n-k+1}^{n^2-n+k}\,c_j\ket{\alpha_j}.
\end{equation}
When we take the difference $\ket{\psi(t)}-\ket{\psi(0)}$, the terms in the eigenspace associated with eigenvalue 1 vanish.

Vectors $\ket{\alpha_j^\pm}$ are self conjugates and $\ket{\psi(0)}$ is real, then Eq.~(\ref{ht_c_j_pm}) implies that $\big|c_j^+\big|^2 = \big|c_j^-\big|^2$. Let us call  $\big|c_j^+\big|^2$ and $\big|c_j^-\big|^2$ by $\big|c_j\big|^2$. Using Eqs.~(\ref{ht_psi_0_generic}) and (\ref{ht_psi_t_generic}), we obtain
\begin{equation}\label{ht_diff}
    \Big\| \ket{\psi(t)}-\ket{\psi(0)} \Big\|^2 = 4 \, \sum_{j=1}^{n-k} \, \big|c_j\big|^2 \, \big(1-T_{2t}(\cos\theta_j)\big),
\end{equation}
where $T_n$ is the $n$-th Chebyshev polynomial of the first kind~\cite{AS}. Using Eq.~(\ref{ht_diff}), we obtain
\begin{equation}
    F(T) = \frac{2}{T+1}\,\sum_{j=1}^{n-k} \, \big|c_j\big|^2 \Big(2\,T+1 - U_{2T}(\cos\theta_j)\Big),
\end{equation}
where $U_n$ is the $n$-th Chebyshev polynomial of the second kind. The quantum hitting time is given by
\begin{equation}
    H_{P,M} = \left\lceil F^{-1}\left(1-\frac{m}{n}\right)\right\rceil.
\end{equation}

\section{Complete Graph}\label{ht_sec_CG}

Let us label the vertices of the complete graph from 1 to $n$ and suppose that the last $m$ vertices are the marked ones. The stochastic matrix of the complete graph is
\begin{equation}
    P = \frac{1}{n-1}\left(n\ket{u^n}\bra{u^n}-I_n\right),
\end{equation}
where $\ket{u^n}=1/\sqrt n \sum_{j=1}^{n} \ket{j}$ is the normalized uniform vector with $n$ components and $\ket{j}$ stands for the $j$-th vector of the computational basis.

Let $P_M$ be the matrix obtained from $P$ by removing the lines and columns corresponding to the marked elements, then
\begin{equation}
    P_M = \frac{1}{n-1}\Big((n-m)\ket{u^{n-m}}\bra{u^{n-m}}-I_{n-m}\Big).
\end{equation}
The characteristic polynomial of $P_M$ is
\begin{equation}
     \left( \lambda-{\frac {n-m-1}{n-1}} \right)  \left( \lambda+ \frac{1}{n
-1} \right) ^{n-m-1}.
\end{equation}
The eigenvector with eigenvalue $(n-m-1)/(n-1)$ is
\begin{equation}\label{ht_nu_n-m_CG}
    \ket{\nu_{n-m}}:=\ket{u^{n-m}}
\end{equation}
and the eigenvectors with eigenvalue $-1/(n-1)$ are
\begin{equation}\label{ht_nu_i_CG}
    \ket{\nu_j}:=\frac{1}{\sqrt{j+1}} \left(\ket{u^j}-\sqrt j\,\ket{j}\right),
\end{equation}
for $1\leq j \leq n-m-1$. That set of eigenvectors forms an orthonormal basis.

The modified stochastic matrix is
\begin{equation}\label{ht_pprime_CG}
    p_{x y}^\prime = \left\{
                       \begin{array}{ll}
                         \frac{1-\delta_{x y}}{n-1}, & \hbox{$1\leq x \leq m-n$;} \\
                         \delta_{x y}, & \hbox{$m-n< x \leq n$.}
                       \end{array}
                     \right.
\end{equation}
All operators of Sec.~\ref{ht_sec_1} must be calculated using the modified matrix. To find the spectral decomposition of $U_{P^\prime}$, the key operator is $C$ given by Eq.~(\ref{ht_C}). The components of $C_{x y}$ are $\sqrt{p_{x y} q_{y x}}$. We have to replace $p_{x y}$ and $q_{x y}$ by $p_{x y}^\prime$.  Using Eq.~(\ref{ht_pprime_CG}) we obtain
\begin{equation}\label{ht_C_CG}
    C = \begin{bmatrix}
  P_M & 0 \\
  0 & I_m \\
\end{bmatrix}.
\end{equation}
$C$ is hermitian, then the nontrivial singular values $ \lambda_j $ are obtained by taking the modulus of the eigenvalues of $ P_M $. The right singular vectors $\ket{\nu_j}$ are the eigenvectors of $P_M $. If the eigenvalue of $ P_M $ is negative, the left singular vector is the negative of the eigenvector of  $ P_M$. These vectors must be increased with $ m $ zeros to have the correct dimension, compatible with the dimension of $C$. Summarizing, $\ket{\nu_{j}}$ and $-\ket{\nu_{j}}$, $1\leq j \leq n-m-1$ are the right and left singular vectors, respectively, with singular value $\cos \theta_1 = \frac{1}{n-1}$, $\ket{\nu_{n-m}}$ is both the right and left singular vectors with singular value $\cos \theta_2 = \frac{n-m-1}{n-1}$. Finally, the submatrix $I_m$ in Eq.~(\ref{ht_C_CG}) adds to the list the singular value 1 with multiplicity $ m $ with the associated singular vectors $\ket{j}$, where $ n-m +1 \leq j \leq n $.

The eigenvectors and eigenvalues of  $U_{P^\prime}$, that can be obtained from the singular values and vectors of $ C $ are given in Table~\ref{ht_table_eigen_CG}. It is missing $ n^2-2n + m $ eigenvectors, all of them associated with eigenvalue 1.

\begin{table}[h]
  \centering
       \begin{tabular}{|c|c|c|}
         \hline
                        &            &  \\
              Eigenvalue & Eigenvector  & Interval\\
                        &            & \\
              \hline
              & & \\
         $\textrm{e}^{\pm2i\theta_1}$ & $\ket{\alpha_j^{\pm}}=\frac{-\big(A+{\textrm e}^{\pm i\theta_1} B\big)\ket{\nu_{j}}}{\sqrt 2 \sin \theta_1}$ & $1\leq j \leq n-m-1$ \\
              & & \\
         $\textrm{e}^{\pm 2i\theta_2}$ & $\ket{\alpha_{n-m}^{\pm}}=\frac{\big(A-{\textrm e}^{\pm i\theta_2} B\big)\ket{\nu_{n-m}}}{\sqrt 2 \sin \theta_2}$ & $j=n-m$ \\
              & & \\
         1 & $\ket{\alpha_j}= A \ket{j}$ & $n-m+1\leq j \leq n$ \\
              & & \\
         \hline
       \end{tabular}
  \caption{Eigenvalues and normalized eigenvectors of $U_{P^\prime}$ obtained from the singular values and vectors of $C$. The vectors $\ket{\nu_{n-m}}$ and $\ket{\nu_{j}}$ are given by Eqs.~(\ref{ht_nu_n-m_CG}) and (\ref{ht_nu_i_CG}) respectively.}\label{ht_table_eigen_CG}
\end{table}

\subsection{Hitting Time on the Complete Graph}

The initial condition in the complete graph reduces to
\begin{equation}\label{ht_ini_cond_CG}
    \ket{\psi(0)} = \frac{1}{\sqrt{n(n-1)}} \sum_{x,y=1}^n (1-{\delta_{xy}}) \ket{x}\ket{y}.
\end{equation}
Using the eigenvectors of Table~\ref{ht_table_eigen_CG}, the expression for $\ket{\psi(0)}$ and the definition (\ref{ht_c_j_pm}), we obtain
\begin{equation}\label{ht_cpm}
    c_j^\pm = \left\{
              \begin{array}{ll}
                0, & \hbox{$1\leq j\leq n-m-1$;} \\
                \frac{\sqrt{n-m}\,\left({1-\textrm{e}^{\mp i\theta_2}}\right)}{\sqrt{2n}\,\,{\sin\theta_2}}, & \hbox{$j=n-m$.}
              \end{array}
            \right.
\end{equation}
where $\theta_2$ is given by
\begin{equation}\label{ht_cos_theta2_CG}
    \cos \theta_2 = \frac{n-m-1}{n-1}.
\end{equation}
The uniform singular vector given by Eq.~(\ref{ht_nu_n-m_CG}) is the only one used in the calculation of the hitting time.

\begin{figure}[h]
\centering
\includegraphics[width=2.5in]{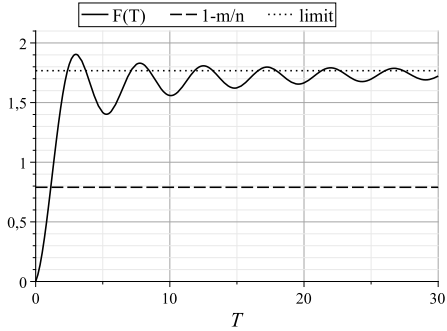}
\caption{Graphs of the function $ F(T) $ (solid line), $1-\frac{m}{n}$ (dashed line) and ${\frac {4 \left( n-1 \right) \left( n-m \right)}{n \left( 2\,n-m-2\right)}}$ (dotted line) for $ n = 100$  and $ m = 21$. The hitting time can be seen in the graph at time $ T $ such that $F(T)=1-\frac{m}{n}$, which is about $1.13$ in this case. } \label{fig:ht_CG}
\end{figure}

The quantity $F(T)$ defined in Eq.~(\ref{ht_D_T}) reduces to
\begin{equation}\label{ht_D_T_CG}
    F(T) = {\frac {2 \left( n-1 \right) \left( n-m \right)  \left(2\,T+1-U_{2T}\left(\frac{n-m-1}{n-1}\right)\right) }{n \left( 2\,n-m-2
 \right)(T+1 )}},
\end{equation}
The graph in Fig.~\ref{fig:ht_CG} shows the behavior of the function $ F(T) $. $ F(T) $ grows rapidly through the dashed line $1-\frac{m}{n}$, then oscillates around the limiting value given by ${\frac {4 \left( n-1 \right) \left( n-m \right)}{n \left( 2\,n-m-2\right)}}$ (dotted line).

For $n\gg m$, the hitting time $ H_{P,M} $ is obtained by employing the method of series inversion on the equation $F(T)=1-\frac{m}{n}$. The first terms are
\begin{equation}\label{ht_H_CG}
    H_{P,M} = \frac{j_0^{-1}\left(\frac{1}{2}\right)}{2}\sqrt{\frac{n}{2\,m}} - \frac{\sqrt{1-\frac{1}{4}\,j_0^{-1}\left(\frac{1}{2}\right)^2}}{1+2\sqrt{1-\frac{1}{4}\,j_0^{-1}\left(\frac{1}{2}\right)^2}} + O\left({\frac{1}{\sqrt n}}\right)
\end{equation}
where $j_0$ is the first spherical Bessel function or the unnormalized sinc function~\cite{AS}. The value of $j_0^{-1}\left(\frac{1}{2}\right)$ is around 1.9.

\subsection{Probability of Finding a Marked Vertex}

The hitting time is used in search algorithms as the running time. It is important to calculate the probability of success at the stopping time. The calculation of the probability of finding a marked element is more elaborated than the calculation of the hitting time, because we have to find $\ket{\psi(t)}$ explicitly, and therefore the eigenvectors with eigenvalue 1 must be considered.

For the complete graph, the eigenvectors that are not orthogonal to the initial condition are $\ket{\alpha_{n-m}^\pm}$ and some of the eigenvectors associated with the eigenvalue 1. Using Eqs.~(\ref{ht_A}) to (\ref{ht_beta_y}) and (\ref{ht_pprime_CG}) we can obtain $\ket{\alpha_{n-m}^{\pm}}$. Substituting $\ket{\alpha_{n-m}^{\pm}}$ and $c_j^\pm$, given by Eq.~(\ref{ht_cpm}), into Eq.~(\ref{ht_psi_t_generic}), we obtain
\begin{eqnarray}\label{ht_psi_t_2_CG}
 & &  \ket{\psi(t)} = \frac{1}{\sqrt{n(n-1)}}\left(\frac {2(n-1)T_{2t}\left(\frac{n-m-1}{n-1}\right)}{2\,n-m-2}\sum_{x,y=1}^{n-m} \big(1-\delta_{x y}\big)\ket{x}\ket{y}\right. + \nonumber \\
 & & \left(\frac {(n-1)T_{2t}\left(\frac{n-m-1}{n-1}\right)}{ 2\,n-m-2}-U_{2t-1}\left(\frac{n-m-1}{n-1}\right)\right)\sum_{x=1}^{n-m}\sum_{y=n-m+1}^n \ket{x}\ket{y}\ + \nonumber \\
& & \left. \left(\frac {(n-1)T_{2t}\left(\frac{n-m-1}{n-1}\right)}{ 2\,n-m-2}+ U_{2t-1}\left(\frac{n-m-1}{n-1}\right)\right)\sum_{x=n-m+1}^n \sum_{y=1}^{n-m} \ket{x}\ket{y} \right)\ + \nonumber \\
& &\sum_{j=n-k+1}^{n^2-n+k}\,c_j\ket{\alpha_j}.
\end{eqnarray} %
The component associated with the eigenvalue 1 can be determined by trial and error directly from the structure of matrix $U_{P^\prime}$. The result is
\begin{eqnarray}\label{ht_eigen_1}
\sum_{j=n-k+1}^{n^2-n+k}\,c_j\ket{\alpha_j} &=& \frac{1}{\sqrt{n(n-1)}} \left(\dfrac{-m}{2n-m-2}\sum_{x,y=1}^{n-m}{\left(1-\delta_{x y}\right)} \ket{x}\ket{y}+\right.\nonumber\\
&&\dfrac{n-m-1}{2n-m-2}\sum_{x=1}^{n-m}\sum_{y=n-m+1}^{n} \big(\ket{x}\ket{y}+\ket{y}\ket{x}\big) +\nonumber\\
&&\left.\sum_{x,y=n-m+1}^{n}{\left(1-\delta_{x y}\right)} \ket{x}\ket{y}\right).
\label{eq:phi1}
\end{eqnarray}

\begin{figure}[h]
\centering
\includegraphics[width=2.5in]{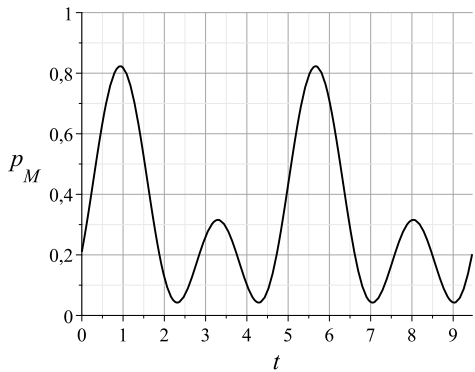}
\caption{Graph of the probability of finding a marked vertex as function of time for $ n = 100$  and $ m = 21$ . The value at $t=0$ is $\frac{m}{n}$ and the function has period $\frac{\pi}{\theta_2}$. } \label{fig:ht_prob_CG}
\end{figure}

The probability of finding a marked element is calculated by using the projector $\cal P_M$ in the vector space spanned by the marked elements, that is
\begin{eqnarray}
  \cal P_M &=& \sum_{x=n-m+1}^n \ket{x}\bra{x}\otimes I_n.
\end{eqnarray}
The probability is given by $\bra{\psi(t)}\cal P_M\ket{\psi(t)}$. Using Eq.~(\ref{ht_psi_t_2_CG}) and (\ref{ht_eigen_1}), we obtain
\begin{eqnarray}
  p_M(t)  &=& \dfrac{m(m-1)}{n(n-1)}+\dfrac{m(n-m)}{n(n-1)}\left(\frac {n-1}{ 2\,n-m-2} T_{2t}\left(\frac{n-m-1}{n-1}\right) + \right.\nonumber\\
&& \left.U_{2t-1}\left(\frac{n-m-1}{n-1}\right)+
\dfrac{n-m-1}{2n-m-2}\right)^{2}. \label{ht_pM_CG}
\end{eqnarray}
The graph of $p_M(t)$ is depicted in Fig.~\ref{fig:ht_prob_CG} when $n=100$ and $m=21$.

The first point of maximum occurs at time
\begin{equation}
t_{\textrm{max}}=
\dfrac{\arctan\left(\dfrac{\sqrt{2n-m-2}}{\sqrt{m}}\right)}{2\arccos\left(\dfrac{n-m-1}{n-1}\right)},
\label{eq:tmax}
\end{equation}
the asymptotic expansion of which is
\begin{equation}
t_{\textrm{max}}=\dfrac{\pi}{4}\sqrt{\dfrac{n}{2\,m}}-\frac{1}{4}+O\left({\frac{1}{\sqrt n}}\right),
\end{equation}
for $n\gg m$. Substituting that result into the expression of probability, we obtain
\begin{equation}
p_{M}(t_{\textrm{max}}) = \dfrac{1}{2}+\sqrt{\dfrac{m}{2\,n}}+O\left({\frac{1}{n}}\right).
\end{equation}
For any values of $ n $ and $ m $, the probability of finding a marked vertex is greater than $\frac{1}{2},$ if the measurement is carried out at time $t_{{\textrm{max}}}$. The instant $t_{{\textrm{max}}}$ is smaller than the hitting time given by Eq.~(\ref{ht_H_CG}), since $\frac{\pi}{4\sqrt 2}\approx 0.56$ while $\frac{j_0^{-1}\left(\frac{1}{2}\right)}{2\sqrt 2}\approx 0.67$. The value of the success probability of an algorithm that uses the hitting time as the running time will be less than the probability at time $t_{\textrm{max}}$. Evaluating $p_M $ at time $ H_{P,M} $ and taking the asymptotic expansion, we obtain
\begin{equation}
p_{M}(H_{P,M}) = \frac{1}{8} \, j_0^{-1}\left(\frac{1}{2}\right)^2 + O\left(\frac{1}{\sqrt n}\right).
\end{equation}
The first term is around $0.45$  and is independent of $ n $ or $ m $. This shows that the hitting time is a good parameter for the stopping point of searching algorithms on the complete graph.


\section*{Acknowledgments}

We acknowledge fruitful discussions with F.~Marquezino and D.~Santiago. R.A.M.S. acknowledges a CAPES' fellowship and R.P. acknowledges CNPq's grant n. 306024/\-2008.


\begin{thebibliography}{10}

\bibitem{MR95}
R.~Motwani and P.~Raghavan, Randomized Algorithms, Cambridge University Press, 1995.


\bibitem{Aldous}
D.~Aldous and J.~Fill, Reversible Markov Chains and Random Walks on Graphs, monograph http://www.stat.berkeley.edu/$\sim$aldous/RWG/ book.html.

\bibitem{Kempe} J.~Kempe, Discrete Quantum Walks Hit Exponentially Faster, RANDOM-APPROX 2003: 354--369
and quant-ph/0205083.

\bibitem{KB06}
H.~Krovi and T.A.~Brun, Quantum walks with infinite hitting times, Phy.~Rev.~A \textbf{74}, 042334 (2006).

\bibitem{KP09}
A.~Kempf and R.~Portugal, Group velocity of discrete-time quantum walks, Phy.~Rev.~A \textbf{79}, 052317 (2009).

\bibitem{Amb03}
A.~Ambainis, {Quantum walk algorithm for element distinctness}, FOCS 2004: 22--31.

\bibitem{Sze04}
M.~Szegedy: Quantum Speed-Up of Markov Chain Based Algorithms. FOCS 2004: 32--41.

\bibitem{MNRS}
F.~Magniez, A.~Nayak, P.C.~Richter and M.~Santha, On the hitting times of quantum versus random walks,
{SODA '09: Proceedings of the Nineteenth Annual ACM -SIAM Symposium on Discrete Algorithms},
86--95, 2009.

\bibitem{Tulsi08}
A.~Tulsi, {Faster quantum walk algorithm for the two dimensional spatial
  search}, Phys. Rev. A \textbf{78}, 012310 (2008).

\bibitem{NC00}
M.A.~Nielsen and I.L.~Chuang, Quantum Computation and Quantum Information, Cambridge University Press, 2000.

\bibitem{AS}
M.~Abramowitz and I.A.~Stegun, Handbook of Mathematical Functions with Formulas, Graphs, and Mathematical Tables, New York: Dover Publications, 1972.


\end{thebibliography}
\end{document}